# Dynamics of magnetic topological solitons in soft magnetic nanostripes

Konstantin Yu. Guslienko, Jun-Young Lee, and Sang-Koog Kim*

*Research Center for Spin Dynamics & Spin-Wave Devices, Seoul National University, Seoul 151-744, South Korea*

The motions of domain walls driven by magnetic field in soft magnetic nanostripes were calculated. The domain walls reveal steady motions in the low fields and oscillations of their internal structure above a critical field. A developed soliton model of the walls explains its dynamics by the motions of a limited number of magnetic topological solitons such as vortex and antivortex. We predict the reduced wall velocity and critical field in the low-field regime, and increased wall-oscillation frequency in nanostripes, compared to 1D Walker solution for bulk magnets. The critical field and velocity are determined by the nanostripe sizes, whereas the oscillation frequency depends only on the field strength.

Domain walls (DWs) in ordered magnetic materials are ten- or hundred-nanometer size transition regions in which the orientation of local magnetization (**M**) gradually changes between neighboring domains of different **M** orientations [1-3]. Recently the motion of DWs driven by magnetic field (or spin polarized current) in geometrically confined magnetic elements such as nanostripes and nanorings became of considerable interest because it is essential to the performance of information storage [4] and logic devices [5]. One of the central issues in this research area is to understand DW motions under an applied magnetic field ($H$) stronger than a certain critical field (the Walker field), $H_w$ [6]. When an external static field **H**=($H$,0,0) is applied to a "head-to head" ("tail-to tail") DW [7] in nanostripes (Fig. 1), the DW propagates along (against) **H** direction to reduce the Zeeman energy [8-10] increasing its velocity with increasing $H$ below $H_w$. But for $H > H_w$, the average velocity $\bar{\upsilon}$ decreases [10]. Such a considerable drop of $\bar{\upsilon}$ in nanostripes was presumably associated with an oscillatory motion of DW [11,12], in analogy to one-dimensional (1D) DW behavior [13, 14]. Oscillatory DW dynamics can be only qualitatively explained within the 1D model [13]. The simulations of DW motion in 2D systems such as nanostripes [15] show that it is also periodic above $H_w$, but the DW internal states are different from the predictions of Ref. 13. Recent experiments have demonstrated the complexities of DW dynamics [8-12,16,17] and different $H_w$ values depending on the nanostripe sizes [11,12]. Static **M** patterns in thin elements can be described within *XY* model via solutions of (1+1) Sine-Gordon equation [18]. However, this approach is not applicable to DW dynamics, where out-of-plane **M** component is crucial. A generalization of 1D-model including the stripe magnetostatics [11,19] cannot explain experiments. Accounting only steady DW motions [20] is not enough to describe DW oscillations. Understanding the DW motions in nanostripes is of special importance in the light of the recent experiments on current driven DW motion [12,16,17]. Well defined oscillations of the stripe resistance [17] need an adequate physical description.

In this Letter, we report an unified approach to the description of moving DWs in nanostripes that reveal periodic transformations in high-$H$ regime as well as a steady movement in low-$H$ regime. The DWs have rich internal structures related to gyrotropic motion of non-linear excitations - magnetic topological solitons [vortices (V) and antivortices (AV)]. The DW dynamics are described as emission, motion and absorption of several magnetic solitons with integer (V and AV) and fractional topological charges. We formulate the DW dynamics in terms of motion of the soliton core positions ($\mathbf{X}_j$) representing the collective dynamic variables and explain how the soliton motion determines the DW transformations and in turn affects the motion of the wall as whole.

The vortex or antivortex walls (VW or AVW) contain a single V (AV) inside the DW structures and also some solitons at the stripe edges [15]. The AV and V bear topological charges $q$ [21] related to their cores (areas where the $\mathbf{M}$ component $M_z \neq 0$) and have non-zero gyrovectors [22]. The gyrovectors result in V (AV) non-linear gyrotropic motion in potential profiles determined by the total magnetic energy $W(\mathbf{X}) = W_m(\mathbf{X}) + W_{ex}(\mathbf{X}) + W_Z(\mathbf{X})$, where $W_m$, $W_{ex}$, $W_Z$ are the magnetostatic, exchange, and Zeeman energies. The periodic dynamic changes of the internal DW structures at $H > H_w$ can be described in terms of the nucleation and annihilation of V or AV at the nanostripe edges [15], that is, the emission and absorption of the magnetic solitons. The transverse walls (TW) have non-zero average transverse $\mathbf{M}$ along $Oy$ axis, directed "upward" (TW$_V$) or "downward" (TW$_A$). These configurations correspond to half-integer ($q = \pm 1/2$, half of the soliton core is inside the nanostripe) topological solitons [15, 23] located on the edges, see Fig. S1 [24]. The VWs (AVWs) have, excepting the charges related to V (AV), also some half-integer charges similar to ones in the TWs [24]. The nucleation of V (AV) changes the edge soliton charges from +1/2 (-1/2) to -1/2 (+1/2). After absorption of V (AV), (-1/2)- or (+1/2)-soliton moves with velocity higher then its counterpart on the other stripe edge creating a distorted TW. At a critical value of the distortion, a new V (AV) will be emitted. The total topological charge of all the solitons inside the nanostripe does not change in course of the DW motion [15]. The sense of the V (AV) gyrotropic motion is determined by the sign of the product $pq$, where $p = +1$ (-1) is the core polarization and $q = +1$ (-1) for V (AV) [21]. V and AV follow the senses of counter-clockwise (CCW)/clockwise (CW) rotation for $qp = +1/(-1)$ [21, 25]. The magnetostatic energy plays an important role in restricted geometry forming a potential well (hill) for V (AV) with respect to the middle of the stripe [15] and the gyromotion is also affected by the sign of the stiffness coefficient $\kappa = |\kappa|\text{sign}(q)$ of the potential $W(\mathbf{X})$. Therefore, the direction of motion of emitted V (AV) depends only on $p$ and they rotate CCW (CW) for $p = +1$ (-1) (see details in Ref. 15). The relatively slow gyromotions lead to negative DW velocities due to V (AV) motions against the **H** direction, and to essential decrease of $\bar{\upsilon}$.

To find the soliton trajectories and traveling times, we consider the Landau-Lifshitz equation of motion accounting $W(\mathbf{X})$ of a soliton in the nanostripe. The dynamics of **M** can be described by the ansatz [22] $\mathbf{M}(\mathbf{r},t) = \mathbf{M}(\mathbf{r} - \mathbf{X}(t))$, where $\mathbf{X}=(X, Y)$ is the soliton (V or AV) core center position in the *x-y* plane, $\mathbf{r}=(x,y)$. This approach is similar to the coordinate-angle approach to DW dynamics by Slonczewski [2], but it is more appropriate for 2D DWs with complex internal structure. The **M** dependence on the coordinate along the nanostripe thickness is neglected because of its small thickness $L$~10 nm. The Lagrangian $\Lambda(\mathbf{X},\dot{\mathbf{X}}) = (\mathbf{G} \times \mathbf{X}) \cdot \dot{\mathbf{X}}/2 - W(\mathbf{X})$ leads to the equation of motion of $\mathbf{X}(t)$, $\mathbf{G} \times \dot{\mathbf{X}} + \hat{D}\dot{\mathbf{X}} - \partial W(\mathbf{X})/\partial \mathbf{X} = 0$, where $\mathbf{G} = -G\hat{\mathbf{z}}$ is the gyrovector, $G = 2\pi M_s L p q / \gamma$, $M_s=|\mathbf{M}|$ and $\hat{D}$ is the damping tensor. **G** is defined via the gyrotensor [22] and is of principal importance for the VW/AVW motion. Assuming decomposition $W_m(\mathbf{X}) + W_{ex}(\mathbf{X}) = const + \kappa Y^2/2$ we get the energy $W_{tot}(\mathbf{X}) = \kappa Y^2/2 - \lambda \mathbf{X} \cdot \mathbf{H} + \mu \hat{\mathbf{z}} \cdot \mathbf{X} \times \mathbf{H}$. The first term describes $W$ variation in the transverse direction of nanostripe ($W$ is invariant along $Ox$ if the stripe is infinitely long), the second and third terms describe the motion of DW as a whole and interaction of V (AV) with **H** [26]. $\kappa$ was calculated for circular dot in Ref. 21, $\lambda = 2M_s wL$ does not depend on the DW internal structure and has sense of the DW magnetic charge per unit of nanostripe length. $\mu$ is approximately given by $\mu = CM_s wL$ for V or $\mu \approx 0$ for AV, where $C=\pm 1$ is the vortex chirality. We rewrite the equation of motion in matrix canonical form using the gyrotensor $\hat{G} = \int dV \hat{g}$, where the gyrodensity is $g_{\alpha\beta} = (M_s/\gamma)\mathbf{m} \cdot [\partial \mathbf{m}/\partial x_\alpha \times \partial \mathbf{m}/\partial x_\beta]$, $\mathbf{m} = \mathbf{M}/M_s$, and the definition $\hat{B} = \hat{G} + \hat{D}$:

$$\dot{\mathbf{X}} = \hat{B}^{-1} \nabla_{\mathbf{X}} W \ . \tag{1}$$

Here $\hat{B}$ components are $B_{ii} = D_{ii}$, $D_{\alpha\beta} = -\alpha_G (M_s/\gamma) \int dV (\partial \mathbf{m}/\partial x_\alpha \cdot \partial \mathbf{m}/\partial x_\beta)$ [22], where $\alpha_G$ is the Gilbert damping parameter, and $G_{xy} = -G_{yx} = G$.

By solving the equations of motion (1) for $X$ and $Y$, the trajectories of V (AV) and corresponding VW (AVW) motions in the nanostripe can be found. Eq. (1) also describes TW motion just assuming $G=0$ and $\partial W/\partial Y = 0$. In the case of any steady motion perpendicularly to the nanostripe length, $\dot{Y} = 0$, Eq. (1) allows quasi-1D solution $\dot{X} = D_{xx}^{-1} \partial W/\partial X$, which corresponds to the well known viscous DW motion derived by Walker [3,13]. This motion is stable in low fields and occurs only for TW or VW in infinitely long stripes (V has $\dot{Y} \neq 0$ and viscous steady motion has no sense for VW for finite $l$, AVW does not exist in such regime). We get using the *XY*-model ansatz [18] $\mathbf{m} = (\cos\Phi, \sin\Phi, 0)$, $\tan\Phi(x,y) = \cos(\pi y/w)/\sinh[\pi(x-X)/w]$, accounting for the half-soliton cores: $D_{xx,yy}^{TW} = -\alpha_G (M_s L/\gamma) I_{xx,yy}^{TW}(\xi)$, $I_{xx,yy}^{TW}(\xi) = 10.03 - \pi \ln\tanh\xi - \pi\xi^2(1/3 \pm 1/6)$, and $D_{xy}^{TW} = 0$, where $\xi = \pi R_c/w$. $D_{\alpha\beta}^{TW}(2D)$ is in 2-2.5 times bigger then $D_{\alpha\beta}^{TW}(1D) = -\alpha_G 2\pi M_s L/\gamma$ [20] calculated from 1D model using $m_y(x) = 1/\cosh[(x-X)/\delta]$, where $\delta$ is the DW width. This leads to the TW velocity in the Walker form $\upsilon_{TW} = \dot{X} = (\lambda/|D_{xx}^{TW}|)H$. We consider VW (AVW) as superposition of V (AV) **M** distribution within the circle of radius $R=w/2$ and TW in the rest of nanostripe [20]. This leads to $D_{\alpha\beta} = D_S \delta_{\alpha\beta} + D_{\alpha\beta}^{TW}$, where the soliton part [26] is $D_S = -\alpha_G \pi M_s L[2 + \ln(R/R_c)]/\gamma$, and $R_c \approx L_{ex} = \sqrt{2A}/M_s$ is the V (AV) core radius. The ratio $|D_{\alpha\beta}/G| \sim \alpha_G \sim 0.01$ is small and $D_{xx}^{VW}/D_{xx}^{TW} \approx 2$ for a typical nanostripe ($w=200$ nm, $R_c=20$ nm). $D_{\alpha\beta}$ for VW and AVW is the same because $D_S$ is determined by the V (AV) exchange energy, which is the same for V and AV [27].

The moving V within the DW experiences a force due to the field $H$, which push the vortex perpendicularly to the stripe length. The equation of motion for $Y$ in Eq. (1) can be written as $\dot{Y} \det(\hat{B}) = \kappa D_{xx}^{VW}(Y - Y_0)$, where $Y_0 = (-\lambda B_{xy}/B_{xx} + \mu)H/\kappa$ is a stationary point for the moving VW.

We assume that the field when $Y_0$ reaches the stripe edge ($Y_0 = \pm w/2$) corresponds to some change of the character of DW motion. This critical field neglecting the small on the ratio ($D/G$) terms $H_w^0 = -\kappa D_{xx}^{VW}/(4M_s L|G|)$ is analog of the Walker field and corresponds to beginning of oscillations of the DW internal structure. $H_W^0$ was called as V "annihilation" field for the quasistatic **M** reversal [28]. Estimation of $\kappa$ similarly to one for a cylindrical dot [21] yields $\kappa = 8\pi M_s^2 L^2/w$ and $H_w^0 = \alpha_G 2\pi M_s (L/w)$ or $H_w^{2D} = H_w^{1D}(L/w)$. I.e., the critical field in nanostripes ($L/w \ll 1$) is essentially reduced in comparison with bulk 1D case. The DW velocity in this field reaches the maximum value $\upsilon_{max} = \kappa w/(2|G|)$. Estimation yields $\upsilon_{max} = 2\gamma M_s L$ for nanostripe that is essentially smaller then the Walker result $\upsilon_{max} = 2\pi\gamma M_s \delta$, if even $\delta$ is reduced due to transverse anisotropy [11]. But in the real nanostripe of finite length the condition $\dot{Y} = 0$ is not satisfied due to weak dependence of $W(\mathbf{X})$ on $X$. V moves up to the stripe edge, annihilates and then VW becomes TW. I.e., the DW steady state motion at $H<H_w$ is moving TW, even if the stable static DW is VW. The characteristic equation of the system of linear differential equations (1) has two eigenvalues $k_1 = 0$ and $k_2 = -\omega_d$, where $\omega_d = |\kappa B_{xx}|/\det(\hat{B}) > 0$ is the relaxation frequency. There are no imaginary eigenvalues, which correspond to V(AV) oscillations, due to absence of confinement potential along the nanostripe length $Ox$. Therefore, Eq. (2) allows general solution in the form:

$$X(t) = X(0) - \frac{\lambda H}{B_{xx}}t + \frac{B_{xy}}{B_{xx}}\left[\frac{w}{2}\frac{H}{H_w} - Y(0)\right](e^{-\omega_d t} - 1), \quad Y(t) = Y(0) + \left[\frac{w}{2}\frac{H}{H_w} - Y(0)\right](1 - e^{-\omega_d t}), \quad (2)$$

where $\mathbf{X}(0) = \mathbf{X}(t=0)$ are the initial conditions, and $H_w/H_w^0 = 1 + O(\alpha_G)$ at $\alpha_G \ll 1$. Eq. (2) describes gyrotropic motion of V (AV) within the VW (AVW) for $H > H_w$ between their emission and absorption at the nanostripe edges. The time variable can be excluded giving explicit equation for the V(AV) trajectories within the nanostripe ($h = H/H_w > 1$, $y = 2Y/w$):

$$X = X(0) - \frac{\lambda H}{B_{xx}} t(Y) - \frac{B_{xy}}{B_{xx}} [Y - Y(0)], \qquad t(Y) = \frac{1}{\omega_d} \ln\left[\frac{h - y(0)}{h - y}\right]. \qquad (3)$$

The trajectories are quite complicated, but for $\alpha_G \ll 1$ we get the simple parabolas $X = (\kappa/2\lambda H)(Y - Y(0))^2 + [\kappa Y(0)/\lambda H - \mu/\lambda](Y - Y(0)) + X(0)$ with $\kappa > 0$ ($\kappa < 0$) for V (AV). $\kappa < 0$ describes instability of AVW at $H=0$. The values of $\kappa$ are different for V (AV) and were calculated numerically. V (AV) trajectory (3) $X'(Y)$ is straight line in the moving frame $X' = X - (\lambda H/|D_{xx}|)t$, i.e. moving V (AV) stays on the line connecting the moving edge solitons. The traveling time $t_S$ of the V (AV) from one stripe edge to the opposite edge can be found from Eq. (3). Let assume that V (AV) was emitted from the edge $y = -w/2$ at $t=0$ and then traveled along trajectory (2) to be adsorbed at $t = t_S$. Putting $Y(0) = -w/2$, $Y = w/2$ we get $t_S = 2\omega_d^{-1} \mathrm{ar\,coth}(h)$ or $t_S = \pi/\gamma H$ within the limit $\alpha_G \ll 1$ (or $h \gg 1$). Surprisingly, it does not depend on the nanostripe sizes $w$, $L$ and intrinsic material parameters $M_s$, $\alpha_G$ as well as on the soliton type. The traveling time $t_{TW}$ of the edge solitons (TWs) between the emission and absorption of V (AV) can be neglected comparing to $t_S$. Therefore, the period of the DW oscillations is $T_w^{2D} \approx 2t_S$ and the eigenfrequency is equal to the Larmor frequency, $\omega_H = \gamma H$. We note that $T_w^{2D}$ is essentially shorter than the Walker period [13] $T_w^{1D} = 2\pi(1 + \alpha^2)/\gamma\sqrt{H^2 - H_w^2}$ just above $H_w$. The DW oscillations for 2D case in restricted geometry are considerably faster than ones for 1D case in bulk materials. The model reveals only weak logarithmic singularity of $T_w^{2D} = -2\omega_d^{-1} \ln(h - 1)$ near $H_w$ in comparison to the power singularity of $T_w^{1D} \sim (h - 1)^{-1/2}$. The increase of the period near the point $h \to 1 + 0$ can be interpreted as critical slow down. To calculate average DW velocity at $h>1$, we consider the ratio $\bar{\upsilon} = (\Delta X_{TW} + \Delta X_S)/(t_{TW} + t_S)$ determining the velocity averaged over the oscillation period, where $\Delta X_{TW}$, $\Delta X_S$ are changes of the positions of TW or soliton DW during the half-period. The slow V (AV) gyrotropic motion corresponds to $\Delta X_S \ll \Delta X_{TW}$, and $t_S \gg t_{TW}$. Therefore, $\bar{\upsilon} \approx \Delta X_{TW}/t_S$, or assuming TW linear motion (as in low-$H$ regime) during the time between V(AV) emission and annihilation it is $\bar{\upsilon} \approx \upsilon_{TW}(t_{TW}/t_S)$. This equation describes a considerable $\bar{\upsilon}$ drop just above $H_w$, $\bar{\upsilon} \cong (0.1 \div 0.2)\upsilon_{max}$ (Fig. 2), which was ob-served experimentally in nanostripes [10]. The mobility ($d\bar{\upsilon}/dH$) of TW $\mu_{TW} = (\lambda/|D_{xx}^{TW}|) = (2\gamma w/\alpha_G)/I_{xx}^{TW}$ for Py stripe with $w$=600 nm [10] is 16.3 m/(s Oe), and $\upsilon_{max} = \mu_{TW} H_w \approx 73.2$ m/s for $H_w$= 4.5 Oe in good agreement with experimental $\upsilon_{max} \approx 75$ m/s.

To confirm analytic calculations and refine the model parameters we conducted micromagnetic simulations [29] assuming rectangular Permalloy (Ni$_{80}$Fe$_{20}$; Py) nanostripes with $L$=10 nm, $l$=6, 12 µm, and $w$=60÷240 nm (see details in Ref. 15). The simulated DW velocity $\bar{\upsilon}$ is plotted vs. $H$ for $w$=140 and 240 nm in Fig. 2. At small $H<H_w$, TW (initially stable VW for $w$=240 nm transforms to TW) moves steadily along the nanostripe. The approximately linear increase of $\bar{\upsilon}$ with increasing $H$ can be described by $\upsilon_{TW} = (\lambda/|D_{xx}^{TW}|)H$. $\bar{\upsilon}$ of TW with $w$=240 nm is in 1.2 times higher than $\bar{\upsilon}$ (140 nm), and is in agreement with analytic calculations (1.5). Just above $H_w$ (here $H_w \approx 10$ Oe), $\bar{\upsilon}$ decreases remarkably [10,11] and the "rigid" DW model fails due to excitations of the DW internal degrees of freedom. Above $H_w$, $\bar{\upsilon}(H)$ is constant, the oscillatory motions of DWs appear [15] as shown in Fig. 2 (inset). The dependences of $t_S = \pi/\gamma H$ for VW and AVW are compared with those obtained from the simulations and are in good agreement (Fig. 3). We also found that $t_{TW}$ is indeed small ($\approx$ 1 ns). Therefore, in high fields (up to ~60 Oe), only VW and AVW represent the DW oscillations. This picture is valid until a chaotic regime revealing multi-soliton states. The simulated V(AV) core trajectories (Fig. 3) were fitted to the parabolic equation of $X(Y)$ and $\kappa$ was found. For the AVW it is negative, whereas for the VW $\kappa$=1.93 erg/cm$^2$ (~2 times bigger then analytic estimation). The simulations yield the values $I_{xx}^{TW} = 9.85$, $I_{xx}^{VW} = 23.90$ close to 12.43 and 26.11 calculated for TW and VW(AVW) analytically. We calculated all the soliton core trajectories (see Fig. S2 [24]) for one of the DW oscillation periods [15], when TW$_V$ transforms first to VW (V emission with *p*=1), second to TW$_\Lambda$ (V absorption), then to VW(*p*=-1), and again to TW$_V$. The Vs (AVs) switch their *p* via their dynamic transformations into the TWs.

Recent experimental observations of DW motion in the nanostripes can be explained by our model. The resistance oscillations due to variable DW position and the DW internal structure transformations were measured by Hayashi *et al*. [17]. The frequencies of these oscillations vary within the range 0.1 ~ 0.5 GHz for the driving fields of 18 ~ 90 Oe above $H_w$ ~14 Oe. The DW of TW$_V$ and TW$_\Lambda$ types have the same resistance, and therefore only half-periods (double eigenfrequencies) of the DW oscillations could be measured applying this technique [17]. The half of the slope $\Delta f / \Delta H$ of the experimental dependences of DW oscillation frequency vs. field in Ref. 17 is equal to 2.7 and 2.8 MHz/Oe, respectively. This value is in good agreement with the slope $\gamma / 2\pi$ =2.9 MHz/Oe for Py according to the equation $1/T_w^{2D} = (\gamma / 2\pi) H$.

The obtained results offer a new physical understanding of the complex DW dynamics in patterned thin films and serve as a non-trivial example of strictly periodic dynamical response of non-linear system to static external perturbations such as DC magnetic fields or currents.

This work was supported by Creative Research Initiatives (Research Center for Spin Dynamics & Spin-Wave Devices) of MOST/KOSEF.

# Figure Captions

**Fig. 1** (color online). The rectangular-shaped nanostripes of thickness $L$=10 nm, length $l$=6 and 12 μm, width $w$, and the coordinate system used. The local magnetization configurations at equilibrium for $w$=140 (a) and 240 nm (b) display a TW with the V-shaped polarization and a VW with the downward core orientation located at the stripe center. An applied static magnetic field $H$ is indicated by the pink arrow. The color wheel indicates the direction of the local magnetization. .

**Fig. 2** (color online). Velocities of domain walls versus $H$ for the stripe width $w$=140 (blue) and 240 nm (orange), $l$=12 μm. The insets show the representative DW-position-vs.-time curves at $H$= 25 Oe for $w$=140 and 240 nm. The velocities were simulated for steady (TW) state at $H<H_w$ or averaged over the wall oscillation period at $H>H_w$.

**Fig. 3** (color online). (a) The trajectories $X(Y)$ of motion of the V (upper plot) and AV cores within the VWs and AVWs for $H$ =15, 20, 25, and 30 Oe. $w$ = 140 nm (upper plot) and 240 nm. The simulated core trajectories (open circles) are well fitted to the analytically derived parabolas (solid lines) described in the text. (b) Traveling time $t_S$ of AV and V from one stripe edge to the opposite edge as function of $H$. Solid line indicates the calculated values of $t_S$ using the equation $t_S = \pi/\gamma H$.

**Figures**

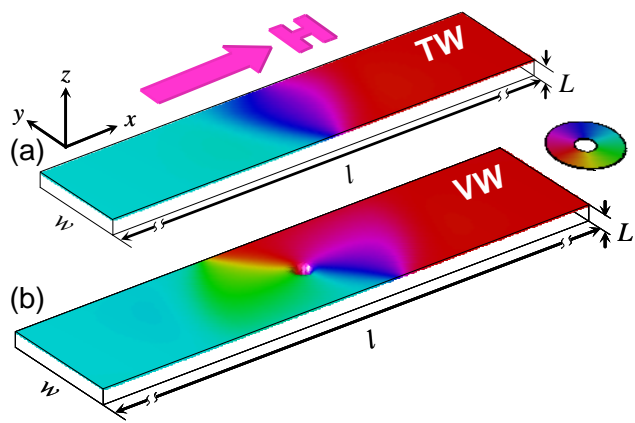

Figure 1

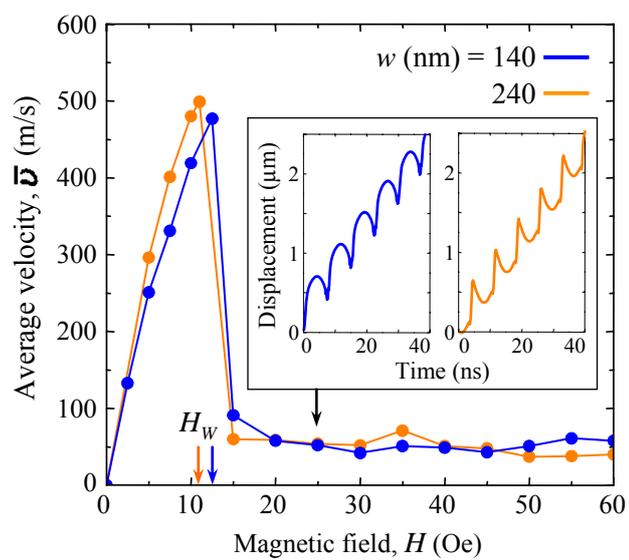

Figure 2

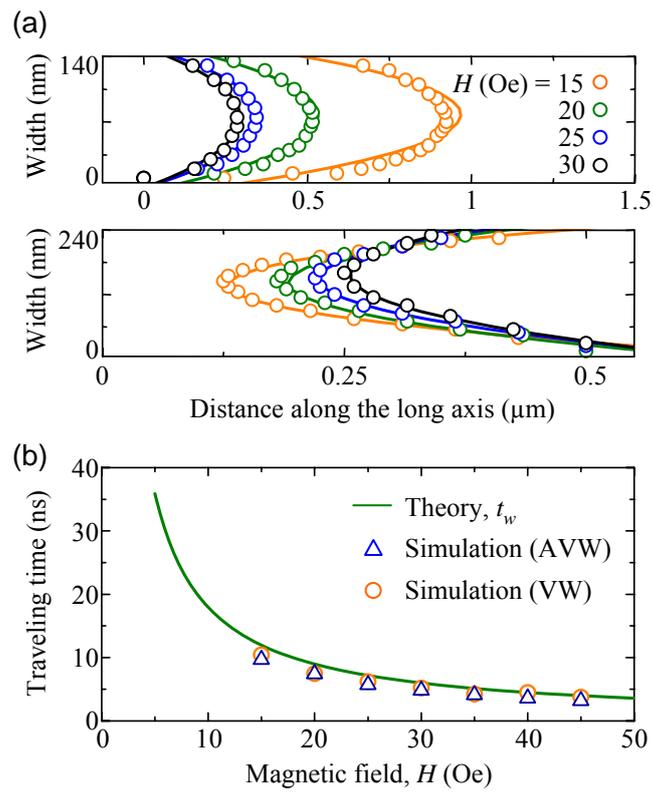

Figure 3